\PassOptionsToPackage{sort&compress,merge,numbers}{natbib}
\documentclass[english]{aipproc}
\usepackage[T1]{fontenc}
\usepackage[latin9]{inputenc}
\usepackage{geometry}
\usepackage{graphicx}

\makeatletter
\layoutstyle{6x9}
\makeatother

\usepackage{babel}
\begin{document}
\title[Joints and quantum cognition]{Joint probabilities and quantum cognition}

\author{J. Acacio de Barros}{   address={Liberal Studies, 1600 Holloway Ave., San Francisco State University, San Francisco, CA 94132},  ,email= {barros@sfsu.edu}}



\begin{abstract}
In this paper we discuss the existence of joint probability distributions for quantum-like response computations in the brain. We do so by focusing on a contextual neural-oscillator model shown to reproduce the main features of behavioral stimulus-response theory. We then exhibit a simple example of contextual random variables not having a joint probability distribution, and describe how such variables can be obtained from neural oscillators, but not from a quantum observable algebra. 
\end{abstract}

\maketitle

\section{Introduction}

In recent years there has been a growing interest in the use of the
concepts and mathematical apparatus of quantum mechanics in the social and behavioral
sciences (see \cite{aerts_quantum_2009,bruza_introduction_2009,pothos_quantum_2009,asano_quantum-like_2010,busemeyer_quantum_2006,busemeyer_empirical_2009,busemeyer_quantum_2007,busemeyer_quantum_2012,de_barros_quantum_2009,de_barros_quantum-like_2012,haven_discussion_2002,haven_black-scholes_2003,haven_wave-equivalent_2004,haven_pilot-wave_2005,khrennikov_importance_2007,khrennikov_quantum_2009,khrennikov_ubiquitous_2010,khrennikova_quantum-like_2012}
and references therein). Most of this work, contrary to Penrose's well 
known idea that brain processes
are actually quantum \cite{penrose_emperors_1989}, proposes that 
social phenomena are better described mathematically by a \emph{quantum-like}
dynamics given by the evolution of a state vector in a Hilbert space, without 
committing to an underlying model that determines
such dynamics. The quantum-like approach has been shown to better fit
empirical data in a variety of experimental conditions, as, for example,
cognitive decision-making processes \cite{asano_quantum-like_2010,busemeyer_quantum_2006,busemeyer_empirical_2009,khrennikov_quantum-like_2007,khrennikov_quantum-like_2009}\footnote{In fact, as Andrei Khrennikov's book-title states, quantum-like
dynamics seems ubiquitous in the social sciences \cite{khrennikov_ubiquitous_2010}.}.

An important question to be understood is why is the quantum formalism 
successful when applied to processes that seem blatantly 
classical. Perhaps
the complex interaction of different classical systems may lead to
the quantum interference of event probabilities 
\cite{de_barros_quantum_2009,de_barros_quantum-like_2012,suppes_phase-oscillator_2012}.
So, if we were to understand the underlying dynamics and answer the
above question, a joint probability distribution and associated joint
expectations of all the random variables corresponding to the observables
being modeled would be a powerful tool. However, since quantum interference
leads to a violation of Kolmogorov's axioms, non-classical quantum-like
processes result in the impossibility of assigning a proper joint
probability distribution to all the random variables corresponding
to observables in the dynamics \cite{de_barros_inequalities_2000,de_barros_probabilistic_2001,de_barros_probabilistic_2010,khrennikov_ubiquitous_2010}. 

We approach the question of how quantum like processes
emerge and what their consequences are by focusing on quantum-cognition. 
We organize
this paper the following way. First, 
following reference \cite{de_barros_quantum_2009}, we briefly argue that 
quantum-like effects in the brain are mainly contextual. 
Then, in the next section, we attempt to build some intuition about the 
origins of quantum-like effects by 
examining a model of brain computations of behavioral responses. 
There, we follow  \cite{suppes_phase-oscillator_2012}, 
where a neural oscillator model 
grounded on reasonable neurophysiological assumptions was developed, 
and then shown in  \cite{de_barros_quantum-like_2012} to
present quantum-like features. Finally, in the last section, we use this model to show an 
example of a possible neural oscillator setup that is contextual, and 
therefore does not have a joint probability distribution for all 
behavioral observables, but that is less restrictive than what is imposed
on the algebra of observables by a Hilbert space structure. We end the paper 
with some discussions about joint probabilities and quantum-like
behavior.

\section{What is quantum in the brain?\label{sec:What-is-quantum}}

Researchers proposing the use of quantum mechanics in the 
brain usually hold two distinct points of view: either the brain is truly performing
quantum computations (see \cite{suppes_quantum_2007} for further details 
and references),
or it is actually a classical system which is better described by the mathematical formalism 
of quantum mechanics. In this section, we argue for the latter. A more
detailed discussion can be found in \cite{de_barros_quantum_2009}. 

In our discussion, we need to make clear what are the differences
between quantum and classical mechanics. Simply speaking,
quantum mechanics puzzled its founders because it departed from classical
mechanics in three main aspects: nondeterminism, contextuality, and
nonlocality. So, let us discuss each of them separately. 

Let us start with nondeterminism. Very early on, Rutherford noticed
that the process of radioactive decay implied a memoryless dynamics,
where the time of decay was not determined by the state of the system
at an earlier time \cite{pais_inward_1986}. As the behavior of quantum
objects became clearer, this nondeterministic behavior seemed like
the norm, and not the exception. Thus, it became clear that the underlying
processes of quantum mechanics were not as in classical mechanics, where
the state of the system at time $t$ completely determined its state
at time $t'>t$. We will not discuss nondeterminism at length,
but instead  make two main general points. 

Our first general point is that, unknown in detail to the
founders of quantum mechanics, a distinction must be made between
determinism and predictability. It is possible for a dynamical system
to be deterministic yet completely unpredictable, to the point where
it is impossible to distinguish it from a purely stochastic system.
Therefore, just because a system shows stochasticity, such as the
radioactive decay, it does not mean that the underlying dynamics is
stochastic. This point is discussed in more details in \cite{weingartner_photons_1996}
(for a different yet complementary view, see \cite{werndl_are_2009}).

Our second point is that, contrary to classical physics, stochastic
mathematical descriptions in social and behavioral sciences are the norm, not the
exception. In the social sciences, as well as cognitive models, stochasticity is seen 
as coming from the description of tremendously complex systems whose details
cannot be known, or even from inherently stochastic laws.
In fact, most non-quantum descriptions of brain processes are stochastic
at some level, and not deterministic. Thus,  deterministic versus
nondeterministic model considerations are not relevant to the macroscopic
description of the brain at the behavioral level, and therefore are
not what really distinguishes quantum-like models from classical models. 

We now turn to contextuality. Early on, physicists noticed that one
of the consequences of the wave description of particles, with momentum
given by the wavelength according to de Broglie's theory, was the
impossibility of describing a system with a coordinate on
the phase space of position and momentum. This led to Heisenberg's
uncertainty principle and to the principle of complementarity. According
to the standard interpretation of quantum mechanics, if two observables
$\hat{A}$ and $\hat{B}$ in a Hilbert space ${\cal H}$ do not commute,
i.e. $[\hat{A},\hat{B}]\neq0$, then a measurement $\hat{A}$ ``disturbs''
the system in such a way that nothing can be said about the values
of $\hat{B}$, unless we measure $\hat{B}$, which then disturbs $\hat{A}$,
and so on. This characteristic of quantum systems is called contextuality 
because the act of measuring the system changes the context in which
the dynamical variables of the system are defined in such a way that
we can no longer say anything about such variables. 
Such contextuality is a large departure from
the classical description of a particle, where any dynamical variable could, in principle, 
be measured simultaneously with as much precision as  desired, without
depending on the context. For example, in classical mechanics momentum,
$p$, and position, $q$, completely determine the state of a system,
and can be used to completely determine the
state of the system at a later time. But in quantum mechanics, the
momentum observable $\hat{P}$ does not commute with the position
observable $\hat{Q}$, and therefore it is not possible to know them
simultaneously, nor to use them to predict exactly the values of both 
$\hat{P}$ and $\hat{Q}$ in a future time. 

But the simultaneous measurement of two variables is not the main issue. 
In fact, contextuality in quantum systems
has even deeper implications. For example, Kochen and Specker showed that the
algebra of observables in a Hilbert space is such that it is impossible
to define values for physical properties of the system that are context
independent \cite{kochen_problem_1975}. 
So, in the case of momentum and position, this result
could be, as it has been by many authors, interpreted as saying that
we cannot consistently assign values for $\hat{P}$ when we know with
exactitude the values of $\hat{Q}$. Because it is not possible to
assign values to variables in a noncontextual way, we can use joint
probabilities to define contextuality the following way: a set of
random variables $\mathbf{X}_{1},\ldots,\mathbf{X}_{n}$ are contextual
if and only if there is no joint probability distribution consistent
with all the marginal distributions observed experimentally (see \cite{suppes_when_1981,suppes_collection_1996,de_barros_probabilistic_2010}). 

We make three remarks about contextuality. First, we emphasize
that contextuality exists in classical physics. For example, classical
fields are contextual, as the solution to field equations depends
on the boundary conditions (i.e., context). In fact, it is possible
to prove that classical fields violate inequalities required for the
existence of non-contextual values, and, more importantly in the case
of continuous variables, that there are observable quantities associated
to classical fields that are incompatible with the existence of a
joint probability distribution \cite{suppes_proposed_1996,suppes_violation_1996}.
This result, of course, does not make contextuality in quantum systems
less puzzling; it just shows how strange it is to refer to
localized particles and at the same time have them behave as fields 
or waves. More importantly, it shows that wave interference
is enough to create contextual variables. 

Our second remark is that
contextuality is common in the social sciences. For example, it is
well known to any social scientist designing questionnaires that
the order of questions is very important. This is so because when
asking a specific question, the context is changed. For example, a
questionnaire on the use of lethal force by the police would get different
answers depending on whether the first questions were about 
abusive behavior by law
enforcement officers or about criminal
activities endangering citizens' lives. 

Finally, our third and
last remark is this. Because classical-field interference can result
in contextual behavior, it is possible to imagine that very complex
systems have similar types of interactions that
can cause such behavior. The brain, certainly a complex system, 
does have wave-like cortical propagations, akin to classical fields 
\cite{robinson_interrelating_2012}.
Thus, contextual
quantum-like behavior is probably happening in the brain. 

Nonlocality is arguably the most puzzling aspect of quantum mechanics,
and it is intimately related to contextuality. In fact, nonlocality
is nothing but "contextuality at-a-distance." As we said above,
a set of random variables is contextual if we cannot provide a joint
probability distribution to them, which is equivalent to say that
we cannot assign values to such variables that are consistent with
all the experimentally observed marginal distributions. This means
that the values of a certain random variable should somehow change
given the other random variables, or context. What is strange in nonlocal
experiments in quantum mechanics is that such contexts seem to instantaneously change
 the values of a variable situated in a far away place, in
a way that appears to be at odds with special relativity%
\footnote{There are some local models that take advantage of some loopholes
or redefinitions of the concept of particle, but their discussion
goes beyond the scope of this paper. For two different approaches,
interested readers should refer to \cite{khrennikov_bells_2007,khrennikov_can_2009,%
suppes_diffraction_1994,suppes_particle_1996,suppes_random-walk_1994,%
suppes_violation_1996,weingartner_photons_1996}. %
}. 

Could we observe nonlocal effects in the brain? To observe nonlocality
we need to make sure that measurements of correlated observables happen
quickly enough to guarantee that they are separated
by spacelike intervals. 
This means that nonlocality in the brain
corresponds to observing correlated processes happening within
a time window of the order of $10^{-10}$ seconds. 
If we cannot observe spacelike 
separation, we could devise nonsuperluminal
mechanisms accounting for the correlations (such as classical fields). 
As far as we know,  no brain decision making processes
occurs in such a small time scale, and therefore 
nonlocal-effects should be  irrelevant in brain or cognitive modeling. 

So, to summarize, in this section we discussed three features that are
considered by some researchers essential to quantum mechanics: contextuality, nonlocality,
and nondeterminism. We argued that among those three, only contextuality
and nondeterminism should be relevant to brain computations, and therefore
to quantum-like effects measured by behavioral 
experiments.

\section{A brief overview of oscillator computations\label{sec:A-brief-overview}}

In this section we present a very schematic description of an oscillator
dynamics that allows for the modeling of behavioral response computation.
Here we only give an intuitive idea of how the model works, and its
underlying motivation. Reference \cite{suppes_phase-oscillator_2012}
gives more details, and the interested reader is directed to it.

There are many different ways to approach how the brain computes responses.
We can try to model neurons all the way down to the synaptic details,
making sure that the action potentials are carefully computed. We
can look at assemblies of neurons, and pay attention to nothing but
coupling strengths and timing of firings. Regardless of how we do this, 
the problem with such detailed
approaches is the need to computationally simulate tens of thousands
of coupled neurons if we want to understand higher functions, such
as language or cognitive response computations. But when doing so,
because of the complexity of the problem, we end up gaining limited
insight into how the brain actually computes responses. To circumvent
this problem, in our previous works  we proposed the use of neural oscillators as a way to
reduce the complexity of the problem while maintaining most of the 
behavioral measurable features \cite{vassilieva_learning_2011,%
suppes_phase-oscillator_2012}. 

In \cite{suppes_phase-oscillator_2012}, neural oscillators 
were used to model the behavioral
stimulus-response theory (SR). 
In SR theory, a trial has the following
structure. First, the trial starts with a set of stimuli with a certain
state of conditioning. Then, a stimulus $s_{n}$ is sampled, and from
it a response is computed. After the response, reinforcement  
occurs, and with probability $c$ the sampled stimulus
is conditioned to the reinforced response. So, SR theory, as described,
is essentially a probabilistic learning theory associating stimulus to response,
and it can not only be easily axiomatized, but also makes clear and
testable predictions that fit experimental data well. 

The basic idea of \cite{suppes_phase-oscillator_2012} is that a robust association
between stimulus and response happens as a collection of neurons fire.
Each collection of neurons is an internal representation of the stimulus
and responses; when neurons associated to a stimulus fire, this corresponds
to the (spreading) activation of such collection. Also, because of
spreading activation, when a stimulus starts firing, the response
oscillators also fire, activating. Since the synaptic interaction
of such neurons leads to their relative synchronization (on or off
phase), the overall dynamics of this large set of neurons (in the
tens of thousands, perhaps) can be tremendously simplified by reducing
the degrees of freedom to one per oscillator, corresponding to their
phase. 

The dynamics determined by the interaction between  
the phase oscillators representing 
stimulus and response can
be described by Kuramoto equations, at least as a first approximation
\cite{izhikevich_dynamical_2007}.
We are then left with the following picture. Once a (distal) stimulus
is presented, through the perceptual system, firing action potentials
activate the neural representation of the stimulus in the brain. Those
neurons start to fire synchronously, and then through spreading activation,
also synchronize with the response oscillators. Such synchronization
may or may not be in phase. Depending on the relative phase, 
a response is selected over the other. In more detail, 
let $s(t)$, $r_{1}(t)$, and $r_{2}(t)$ represent stimulus and
response oscillators, respectively, and let us assume for simplicity
that they are given by 
\begin{eqnarray}
s_{i}(t) & =A\cos\left(\varphi_{s_{i}}(t)\right)= & A\cos\left(\omega_{0}t\right),\label{eq:oscillation-s}\\
r_{1}(t) & =A\cos\left(\varphi_{r_{1}}(t)\right)= & A\cos\left(\omega_{0}t+\delta\phi_{1}\right),\label{eq:oscillation-1}\\
r_{2}(t) & =A\cos\left(\varphi_{r_{2}}(t)\right)= & A\cos\left(\omega_{0}t+\delta\phi_{2}\right).\label{eq:oscillation-2}
\end{eqnarray}
The arguments $\varphi_{s}(t)$, $\varphi_{r_{1}}(t)$, and $\varphi_{r_{2}}(t)$
are their phases, and $\delta\phi_{1}$ and $\delta\phi_{2}$ are
constants (all amplitudes $A$ are assumed to be the same). As 
described in \cite{suppes_phase-oscillator_2012}, since neural oscillators
have a wave-like behavior, their dynamics
satisfy the principle of superposition, thus making oscillators prone
to interference effects. As such, the mean intensity in time gives
us a measure of the excitation of the neurons in the neural oscillators.
For the superposition of $s(t)$ and $r_{1}(t)$, 
\begin{eqnarray*}
I_{1} & = & \left\langle \left(s_{i}(t)+r_{1}(t)\right)^{2}\right\rangle _{t},
\end{eqnarray*}
where the subscript $t$ refers to the time average. It is easy
to compute that 
\[
I_{1}=A^{2}\left(1+\cos\left(\delta\phi_{1}\right)\right),\qquad I_{2}=A^{2}\left(1+\cos\left(\delta\phi_{2}\right)\right).
\]
From the above equations, the maximum intensity is $2A^{2}$, and
the minimum is zero. Thus, the maximum difference between $I_{1}$
and $I_{2}$ happens when their relative phases are $\pi$. We also
expect a maximum contrast between $I_{1}$ and $I_{2}$ when a response
is not in between the responses represented by the oscillators $r_{1}\left(t\right)$
and $r_{2}\left(t\right)$. For in between responses, we should expect
less contrast, with the minimum contrast happening when the response
lies on the mid-point of the continuum between the responses associated
to $r_{1}(t)$ and $r_{2}(t)$. 

Intuitively, we can think of the oscillator model 
the following way. When oscillators synchronize with relative
phases such that $I_{1}$ is max and $I_{2}$ is zero,
then clearly the response is toward $r_{1}\left(t\right)$. 
When the
synchronization happens with phases such that both $I_{1}$ and $I_{2}$
are the same, we can think of $I_{1}$ and $I_{2}$ conducting a ``tug-of-war''
between the two responses, with both being equally activated. This balance
of responses happens if we impose
\begin{equation}
\delta\phi_{1}=\delta\phi_{2}+\pi\equiv\delta\phi,\label{eq:ideal-phase-diff}
\end{equation}
which results in 
\begin{equation}
I_{1}=A^{2}\left(1+\cos\left(\delta\phi\right)\right),\label{eq:phase-1}
\end{equation}
 and
\begin{equation}
I_{2}=A^{2}\left(1-\cos\left(\delta\phi\right)\right).\label{eq:phase-2}
\end{equation}
From equations (\ref{eq:phase-1}) and (\ref{eq:phase-2}), let $x\in[-1,1]$
be the normalized difference in intensities between $r_{1}$ and $r_{2}$,
i.e. 
\begin{eqnarray}
x & \equiv & \frac{I_{1}-I_{2}}{I_{1}+I_{2}}=\cos\left(\delta\varphi\right),\label{eq:angle-reinforcement-b}
\end{eqnarray}
$0\leq\delta\varphi\leq\pi$. So, we can use arbitrary phase differences
between oscillators to code for a continuum of responses between $-1$
and $1$. 

We showed in this section that  by coding phase differences between neural oscillators
we can compute a response. Stochasticity comes from assumptions about
initial conditions, learning, and sampling, and contextually comes from 
the interference of neural oscillators.

\section{Joint probabilities and brain computations\label{sec:Joint-probabilities-and}}

From the discussions above we see that interfering neural oscillators
may have quantum-like behavior. In fact,  reference 
\cite{de_barros_quantum-like_2012}
used such model 
to reproduce a violation of Savage's sure-thing principle (STP), a
widely used example to show quantum-like effects in 
psychology. In
this section, we will sketch the arguments of  \cite{de_barros_quantum-like_2012}
and then present a  
more complex neural-oscillator set exhibiting strongly 
contextual correlations.

Savage's STP can be described the following way. Imagine someone
needs to make a decision between $X$ and $\neg X$, but is unsure about 
how $X$ fares under condition $A$. After careful
consideration, she finds out that she would prefer $X$ over $\neg X$
if $A$ is true, and she would also prefer $X$ over $\neg X$ if $A$ is false.
Thus, she should decide for $X$ 
over $\neg X$ regardless of whether
she knows $A$ or $\neg A$ to be true. As Savage stated \cite{savage_foundations_1972}, 
"no other
extralogical principle governing decisions that finds such ready acceptance"
is known, and he called it the sure-thing principle.
This principle can also be shown to be a consequence of
 Kolmogorov's axioms of probability 
\cite{de_barros_quantum-like_2012}, which in turn are shown to
be a consequence of requiring measures of belief to follow the rules
of logic \cite{jaynes_probability_2003}. In other words, rational
 agents should follow STP. 
 
However, as Tversky and Shafir showed,
people often violate STP \cite{tversky_disjunction_1992}. To 
explain such violation, many quantum-like models were proposed 
 (see \cite{khrennikov_ubiquitous_2010} and references therein), 
 as interference in quantum mechanics leads to violations of Kolmogorov's 
 axioms.
As we mentioned above, here we focus on the quantum-like model presented in 
\cite{de_barros_quantum-like_2012}. The basic idea is that for conditions
$A$ and $\neg A$ we associate two neural oscillators $s_{A}$ and
$s_{\neg A}$. If $A$ is known, only $s_{A}$ is activated, resulting
in a given response, say $X$, coded by the synaptic couplings between
$s_{A}$ and $r_{1}$ and $r_{2}$. Similarly, if $\neg A$ is known,
$s_{\neg A}$ is activated, and $X$ is once again computed. However,
when the condition is unknown, both oscillators are activated.
Since $A$ and $\neg A$ are incompatible, in the sense that $A$
is $\neg\left(\neg A\right)$, we should expect in the model a phase difference 
between them, which translates into inhibitory synapses. The inhibitory
synapses  create a phase relation resulting in an interference effect
between the two oscillators, and we can see them as couplings 
corresponding to coherency
relation between the  stimulus and response oscillators. 
Thus, when computing the result with both
oscillators activated, it is possible to prefer
$\neg X$ over $X$, a violation of STP. 

It is instructive to relate the oscillator model
to the two-slit experiment of quantum mechanics. In the two slit
experiment, the probability of detection when both slits are open
is not the sum of the probabilities of the particle going through
one slit or the other (when one of the slits is closed), because of
quantum interference. Another way to think about it is that there
is no joint probability distribution for the wave and particle characteristics
of a quantum particle. So, the two-slit experiment does not satisfy
Kolmogorov's rule of probabilities. Similarly, for the oscillator
model, when only one oscillator is active, we have a certain probability
of response; but when both oscillators are active, the probabilities
do not add as classical probabilities because
of interference. 

Now that we showed how neural oscillators interfere, 
let us turn to a different example where no joint probability
distribution exists. The simplest set of random variables not having a joint probability distribution
is the following. Let $\mathbf{X}$, $\mathbf{Y}$, and $\mathbf{Z}$
be $\pm1$-valued random variables with mean zero, and let $E\left(\mathbf{XY}\right)$,
$E\left(\mathbf{XZ}\right)$, and $E\left(\mathbf{YZ}\right)$ be
measurable and known quantities. Then, it is possible to prove that
there exists a joint probability distribution for $\mathbf{X}$, $\mathbf{Y}$,
and $\mathbf{Z}$ if and only if \cite{suppes_when_1981,suppes_collection_1996}
\[
-1\leq E\left(\mathbf{XY}\right)+E\left(\mathbf{XZ}\right)+E\left(\mathbf{YZ}\right)\leq1+2\min\left\{ E\left(\mathbf{XY}\right),E\left(\mathbf{XZ}\right),E\left(\mathbf{YZ}\right)\right\} .
\]
For example, if
\[
E\left(\mathbf{XY}\right)=E\left(\mathbf{XZ}\right)=E\left(\mathbf{YZ}\right)=-1,
\]
from the inequality above we would have no joint probability distribution. In fact, 
it is easy to see why in this example. Say in a given trial we measure
$\mathbf{X}=1$. Then, from the correlations, it follows that $\mathbf{Y}=-1$,
and then $\mathbf{Z}=1$, and finally $\mathbf{X}=-1$, clearly a
contradiction. Thus, the values of $\mathbf{X}$, $\mathbf{Y}$, 
and $\mathbf{Z}$ cannot be measured simultaneously, as any simultaneous 
measure of all three variables would be inconsistent with the 
marginal distributions. 

This three variable example is relevant to us for two reasons. First, it can
be reproduced as responses computed with neural oscillators. Second, 
though it is highly contextual, it cannot be derived from  quantum 
mechanics. Let us examine each one of those claims separately, starting 
with a neural oscillator model for the correlations. 

\begin{figure}
\begin{centering}
\includegraphics[scale=0.7]{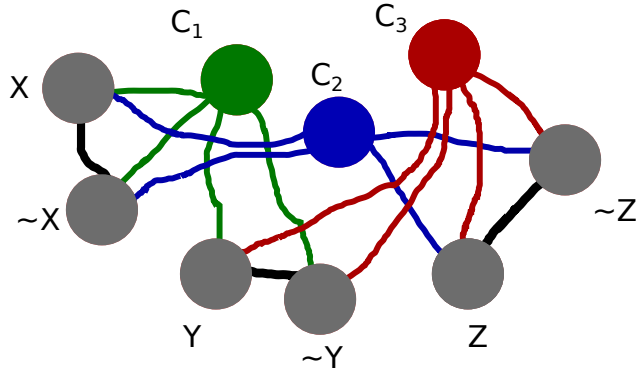}
\par\end{centering}
\caption{\label{fig:Neural-oscillator-model}Neural oscillator model of the
three random variable example. Here the three stimulus oscillators
are represented by $C_{1}$, $C_{2}$, and $C_{3}$, and each stimulus
is coupled to four response oscillators. Response oscillator $X$ 
corresponds to $\mathbf{X}=1$, \textasciitilde$X$ to $\mathbf{X}=-1$, and so on.  }
\end{figure}

The stimulus and response oscillators needed to model the correlations 
are shown in Figure \ref{fig:Neural-oscillator-model}.
It is easy to see how this oscillator model can produce the same
correlations as the three random variables, by having each 
stimulus oscillator be a given context (not necessarily compatible with 
another oscillator's context). We should emphasize that there are no 
matter-of-fact reasons for us to believe that, in the oscillator model, all
three responses $\mathbf{X}$, $\mathbf{Y}$, and $\mathbf{Z}$ cannot
be simultaneously computed. We could indeed imagine a behavioral experiment
where contradictory responses from conditioned contexts could be elicited. 

We now turn to quantum mechanics. 
To measure the correlations shown above, we need to pairwise measure 
each random variable. 
That means that $\mathbf{X}$,
$\mathbf{Y}$, and $\mathbf{Z}$ pairwise commute. Thus, it is possible to
find a basis in the Hilbert space where all corresponding quantum
observables $\hat{X}$, $\hat{Y}$, and $\hat{Z}$ are diagonal, 
making it possible to measure $\mathbf{X}$,
$\mathbf{Y}$, and $\mathbf{Z}$ simultaneously. Since we can measure
then simultaneously, there must exists a joint probability distribution.
Therefore, if a system is quantum mechanical, it cannot present 
correlations that are too strong as to not have a joint probability distribution, 
which rules out $-1$ or even $-1/3$. 

The example shown illustrates an important point. In addition to 
non-Kolmogorovian probabilities, quantum mechanics brings a 
constraint on correlations encoded by the algebra of the Hilbert space.
Furthermore, quantum dynamics requires more than just 
the assumption that physical states are described by vectors
in a Hilbert space \cite{simon_no-signaling_2001}. 
It should be emphasized that 
these constraints are quantum mechanical, and they do not apply 
to the oscillator model, and possibly not to the brain.

\section{Conclusions}

In this paper we discussed a contextual neural-oscillator brain model
based on reasonable neurophysiological assumptions. We sketched how
this neural-oscillator model could not only be used to obtain standard
quantum-like features, such as a violation of Savage's sure-thing
principle, but also how it could be used to encode contextual random
variables with no joint probability
distribution. In fact, the oscillator-based system shown presents correlations
that are so strong as to be incompatible with a Hilbert space 
representation of the corresponding
observables. 

We are then left with the question of how to better represent quantum-like
dynamics in the brain, since Hilbert spaces seem too restrictive.
Arguably, relaxing Kolmogorov's axioms might be a feasible approach.
For example, if we allow for the (unobservable) joint probabilities
to be negative, we could use such values to compute correlations on
unobserved situations (like the $\mathbf{X}$, $\mathbf{Y}$, and
$\mathbf{Z}$ random variables)
and test them against possible experimental conditions. So, we end
with the following additional questions. Would the use of non-standard probabilities,
perhaps guided by intuition from quantum mechanics%
\footnote{An example of such guidance is the clever use of the master equation
in reference \cite{khrennikova_quantum-like_2012}.%
}, be a more appropriate tool to represent quantum-like effects in
the social sciences than the use of vectors on Hilbert spaces? 
What type of formalism could at the same time give the required tools to 
represent contextual quantum-like dynamics without imposing
the unnecessary constraints from a Hilbert space representation?

\bibliographystyle{plainnat}
\bibliography{VaxjoProbPaper}

\end{document}